\documentstyle[psfig]{mn}

\def\etal{{\it et al.\ }}
\def\eg{{\it e.g.\ }}

\def\ie{{\it i.e.\ }}

\def\spose#1{\hbox to 0pt{#1\hss}}
\def\approxlt{\mathrel{\spose{\lower 3pt\hbox{$\sim$}}
	\raise 2.0pt\hbox{$<$}}}
\def\approxgt{\mathrel{\spose{\lower 3pt\hbox{$\sim$}}
	\raise 2.0pt\hbox{$>$}}}
\def\approxpropto{\mathrel{\spose{\lower 3pt\hbox{$\sim$}}
	\raise 2.0pt\hbox{$\propto$}}}
\mathchardef\twiddle="2218

\def\multleft#1{\hbox to size{\vbox {\halign {\lft{##}\cr #1}}\hfill}\par}
\def\multright#1{\hbox to size{\vbox {\halign {\rt{##}\cr #1}}\hfill}\par}

\def\today{\ifcase\month\or January\or February\or March\or April\or May\or
      June\or July\or August\or September\or October\or November\or December\fi
      \space\number\day, \number\year}
\def\<{\thinspace}

\def\erg{{\rm\thinspace erg}}

\def\km{{\rm\thinspace km}}

\def\Mpc{{\rm\thinspace Mpc}}
\def\Msun{\hbox{$\rm\thinspace M_{\odot}$}}

\def\s{{\rm\thinspace s}}

\def\ergps{\hbox{$\erg\s^{-1}\,$}}

\def\kmps{\hbox{$\km\s^{-1}\,$}}

\def\kmpspMpc{\hbox{$\kmps\Mpc^{-1}$}}

\title[The X-ray virial relations for relaxed lensing clusters 
observed with Chandra]
{The X-ray virial relations for relaxed lensing clusters 
observed with Chandra}
\author[S.W. Allen, R.W. Schmidt \& A.C. Fabian]
{\parbox[]{6.in} {S.W. Allen, R.W. Schmidt and A.C. Fabian \\
\footnotesize
Institute of Astronomy, Madingley Road, Cambridge CB3 0HA
 }}
\begin{document}
\maketitle
\begin{abstract}
We examine the relations linking mass, X-ray temperature and
bolometric luminosity for a sample of luminous, relatively relaxed
clusters of galaxies  observed with the Chandra Observatory, for which
independent confirmation of the mass results is available  from
gravitational lensing studies. Within radii corresponding  to a fixed
overdensity $\Delta = 2500$ with respect to the critical density at
the  redshifts of the clusters, the observed temperature profiles,
scaled in  units of $T_{2500}$ and $r_{2500}$, exhibit an
approximately universal  form which rises within $r \sim
0.3\,r_{2500}$ and then remains  approximately constant out to
$r_{2500}$. We obtain best-fit slopes for the mass-temperature and
temperature-luminosity relations consistent with the  predictions from
simple scaling arguments \ie $M_{2500} \propto T_{2500}^{3/2}$ and
$L_{2500} \propto T_{2500}^2$, respectively. We confirm the presence
of a  systematic offset of $\sim 40$ per cent between the
normalizations of the  observed and predicted mass-temperature
relations for both SCDM and  $\Lambda$CDM cosmologies.
\end{abstract}

\begin{keywords}
X-rays: galaxies: clusters -- galaxies: clusters: general --
gravitational lensing -- cosmological parameters

\end{keywords}

\section{Introduction}

The spatial distribution, mass function and redshift evolution of
clusters of  galaxies are sensitive functions of cosmology. The space density 
$n(M,z)$ of clusters predicted by analytical models (\eg Press \& Schechter 
1974; Lacey \& Cole 1993; Sheth, Mo \& Tormen 2001) and numerical
simulations (\eg Navarro, Frenk \& White 1995; Eke, Cole \& Frenk 1996; 
Jenkins \etal 2000; Bode \etal 2001) can be related to (more easily) 
observable properties such as the
X-ray temperatures and  luminosities of clusters via simple scaling
relations. Assuming that the X-ray gas in clusters is virialized and
in hydrostatic equilibrium, the mass, $M_\Delta$, within radius
$r_\Delta$ (inside which  the mean mass density is $\Delta$ times the
critical density, $\rho_c(z)$,  at that epoch) is related to the mean
mass-weighted temperature within that radius, $T_\Delta$, by
$E(z)M_\Delta  \propto T_\Delta^{3/2}$. Here, $E(z) = H(z)/H_0 = 
(1+z)\sqrt{(1+z\Omega_{\rm
m}+\Omega_{\Lambda}/(1+z)^2-\Omega_{\Lambda})}$, where $H(z)$ is the 
redshift-dependent  Hubble Constant (\eg Bryan \& Norman
1998). Since the X-rays from rich clusters are primarily bremsstrahlung 
emission, one can also show that
$L_\Delta/E(z) \propto T_\Delta^2$,  where
$L_\Delta$ is the bolometric luminosity from within radius
$r_\Delta$. The validity of these simple scaling relations is
supported by  numerical simulations (\eg Evrard, Metzler \& Navarro
1996; Bryan \& Norman 1998; Thomas \etal 2001; Mathiesen \& Evrard
2001), although the normalization  of the mass-temperature relation
exhibits some variation from study to study (the normalization of
Bryan \& Norman 1998 is 17 per cent higher than that of Evrard,
Metzler \& Navarro 1996 for $\Delta = 250$; see also Table 1 of 
Afshordi \& Cen 2001).  The normalization of the luminosity-temperature 
relation is more difficult to predict due to the potentially complex physics 
of the X-ray gas in the innermost regions of clusters from where the bulk 
of the X-ray luminosity arises. 

Recent observational determinations of the mass-temperature
relation, based on ASCA and ROSAT data for relatively hot ($kT
\approxgt 3-4$ keV) clusters (\eg Horner, Mushotzky \& Scharf 1999;
Nevalainen, Markevitch \& Forman 2000; Finoguenov, Reiprich  \&
B\"ohringer 2001) have recovered a slope consistent 
with the simple scaling-law predictions, although the observed
normalizations are typically $\sim 40$ per cent lower than predicted by 
the simulations of  Evrard, Metzler \& Navarro (1996) for a 
standard cold dark matter (SCDM) cosmology. For clusters at lower 
temperatures, some steepening of the mass-temperature relation is 
inferred (Nevalainen \etal 2000, Finoguenov \etal 2001).  Studies 
of the luminosity-temperature
relation (\eg White, Jones \& Forman  1997; Allen \& Fabian 1998;
Markevitch 1998; Arnaud \& Evrard 1999) have  generally measured $L_{\rm
Bol} \propto T^3$, whereas theory predicts  $L_{\rm Bol} \propto
T^2$. This has been taken as evidence for significant  pre-heating
and/or cooling in cluster cores (\eg Kaiser 1991; Evrard \&  Henry
1991; Cavaliere, Menci \& Tozzi 1997; Pearce \etal 2000; Bialek, Evrard 
\& Mohr 2001). Allen \& Fabian 
(1998) have shown that for hot ($kT \approxgt 5$ keV), relaxed clusters 
$L_{\rm Bol} \approxpropto T^2$ is recovered once the effects of  cool,
central components are accounted for in the spectral X-ray analysis,
suggesting  (in agreement with the later mass-temperature results) that
pre-heating may only significantly affect the properties of cooler, 
less-luminous clusters.

A major goal of studies with the new generation of X-ray missions including 
the Chandra Observatory and XMM-Newton, which permit the first direct 
spatially-resolved X-ray spectroscopy of hot, distant  clusters, is the 
verification and accurate calibration
of the virial  relations for galaxy clusters. In particular, detailed
studies of systems for  which precise mass measurements have been made
using other, independent  methods are required. An early attempt at
combining X-ray and gravitational  lensing data for clusters observed
with the ASCA satellite to study the  mass-temperature relation was
presented by Hjorth, Oukbir \& van Kampen (1998).  In this letter we
use new Chandra observations to determine the X-ray virial  relations
for a sample of luminous, relatively relaxed clusters
spanning the redshift range $0.1 <z<0.45$, for which lensing mass
measurements  are available and have been shown to be in good
agreement with the Chandra  results (Section 2; see \eg Allen 1998, 
B\"ohringer \etal 1998 for earlier results). We
present gas mass-weighted temperatures, bolometric luminosities and total
mass measurements within radii  corresponding to a fixed overdensity
$\Delta = 2500$ at the redshifts of the clusters, and compare the
observed scaling relations between these quantities with those predicted 
by simulations. Results are given for two cosmologies: SCDM with 
$h = H_0/100$\kmpspMpc $= 0.5$, $\Omega_{\rm m} = 1$  and
$\Omega_\Lambda = 0$, and $\Lambda$CDM with $h=0.7$,
$\Omega_{\rm m} = 0.3$ and $\Omega_\Lambda = 0.7$.

\section{Observations and data analysis}

\begin{table}
\begin{center}
\caption{Summary of the Chandra observations. 
}\label{table:targets}
\vskip -0.2truein
\begin{tabular}{ c c c c c c c }
&&&&  \\
              & ~ &  z  & Date   &  Net Exposure  \\
\hline
PKS0745-191      & ~ &  0.103 & 2001 Jun 16 & 17.9 \\
Abell 2390       & ~ &  0.230 & 1999 Nov 7  & 9.1  \\
Abell 1835       & ~ &  0.252 & 1999 Dec 12 & 19.6 \\
MS2137-2353      & ~ &  0.313 & 1999 Nov 18 & 20.6 \\
RXJ1347-1145(1)  & ~ &  0.451 & 2000 Mar 05 & 8.9 \\
RXJ1347-1145(2)  & ~ &  0.451 & 2000 Apr 29 & 10.0 \\
3C295            & ~ &  0.461 & 1999 Aug 30 & 17.0 \\
&&&& \\		    
\hline			    
\end{tabular}
\end{center}
\end{table}

The Chandra observations were carried out using the back-illuminated S3 
detector on the Advanced CCD Imaging Spectrometer (ACIS) between 1999 
August 30 and 2001 June 16. For our analysis we have used the 
the level-2 event lists provided by the standard Chandra pipeline 
processing. These lists were cleaned for periods of background flaring 
using the CIAO software package resulting in the net exposure times 
summarized in Table \ref{table:targets}. 

The Chandra data have been analysed using the methods described by
Allen \etal (2001b,c) and Schmidt, Allen \& Fabian (2001). In brief, 
concentric annular 
spectra were extracted from the cleaned event lists, centred on the peaks of 
the X-ray emission from the clusters.\footnote{For RXJ1347-1145, the data 
from the southeast quadrant of the cluster were excluded due to 
ongoing merger activity in that region; Allen \etal (2001c).} The spectra 
were analysed using XSPEC (version 11.0: Arnaud 1996), the MEKAL plasma 
emission code (Kaastra \& Mewe 1993; incorporating the Fe-L calculations 
of Liedhal, Osterheld \& Goldstein 1995), and the photoelectric absorption 
models of Balucinska-Church \& McCammon (1992). Two  separate models were
applied to the data, the first of which was fitted to each annular spectrum 
individually in order to measure the projected temperature profiles. The 
second model was applied to all annuli simultaneously, in order to 
determine the deprojected temperature profiles under the assumption 
of spherical symmetry. Only data in the $0.5-7.0$ keV range 
were used.

\begin{table*}
\begin{center}
\caption{The best-fit parameter values and 68 per cent 
($\Delta \chi^2=1.0$) confidence limits for the NFW mass models.
$r_{s}$ values are in units of Mpc and $\sigma$ values in \kmps.}
\label{table:nfw}
\vskip 0.05truein
\begin{tabular}{ c c c c c c c c c }
\multicolumn{1}{c}{} &
\multicolumn{1}{c}{} &
\multicolumn{3}{c}{SCDM} &
\multicolumn{1}{c}{} &
\multicolumn{3}{c}{$\Lambda$CDM} \\
              & ~ &  $r_{s}$                &       c                 &     $\sigma$         & ~ &   $r_{s}$                &       c                 &    $\sigma$        \\       
\hline											              
PKS0745-191   & ~ & $0.85^{+0.12}_{-0.18}$  & $3.66^{+0.53}_{-0.26}$  & $1275^{+75}_{-125}$  & ~ &  $0.64^{+0.09}_{-0.12}$  & $3.83^{+0.52}_{-0.27}$  & $1275^{+75}_{-100}$ \\
Abell 2390    & ~ & $0.79^{+1.38}_{-0.39}$  & $3.28^{+1.77}_{-1.51}$  & $1250^{+600}_{-275}$ & ~ &  $0.76^{+1.59}_{-0.39}$  & $3.20^{+1.79}_{-1.57}$  & $1350^{+775}_{-325}$ \\
Abell 1835    & ~ & $0.64^{+0.21}_{-0.12}$  & $4.02^{+0.54}_{-0.64}$  & $1275^{+150}_{-100}$ & ~ &  $0.55^{+0.18}_{-0.09}$  & $4.21^{+0.53}_{-0.61}$  & $1300^{+175}_{-75}$ \\
MS2137-2353   & ~ & $0.18^{+0.05}_{-0.02}$  & $8.36^{+0.69}_{-1.25}$  & $800^{+70}_{-30}$    & ~ &  $0.16^{+0.03}_{-0.03}$  & $8.71^{+1.22}_{-0.92}$  & $810^{+50}_{-60}$    \\
RXJ1347-1145  & ~ & $0.40^{+0.24}_{-0.12}$  & $5.87^{+1.35}_{-1.44}$  & $1450^{+300}_{-200}$ & ~ &  $0.37^{+0.18}_{-0.12}$  & $6.34^{+1.61}_{-1.36}$  & $1475^{+250}_{-225}$ \\
3C295         & ~ & $0.19^{+0.07}_{-0.05}$  & $6.92^{+1.67}_{-1.37}$  & $820^{+90}_{-80}$    & ~ &  $0.16^{+0.07}_{-0.04}$  & $7.90^{+1.71}_{-1.72}$  & $800^{+100}_{-80}$    \\
&&&&&&&& \\
\hline
\end{tabular}
\end{center}
 \parbox {7in}
{}
\end{table*}

For the mass modelling, azimuthally-averaged surface brightness
profiles  were constructed from background subtracted, flat-fielded
images with a  $0.984\times0.984$ arcsec$^2$ pixel scale ($2\times2$
raw detector  pixels). When combined with the deprojected spectral
temperature profiles,  the surface brightness profiles can be used to
determine the X-ray gas mass  and total mass profiles in the
clusters. For this analysis we have used an enhanced version of the
image deprojection code described by White,  Jones \& Forman
(1997)\footnote{The observed surface brightness  profile and a
particular parameterized mass model are together used to  predict the
temperature profile of the X-ray gas. (We use the median  temperature
profile determined from 100 Monte-Carlo simulations. The  outermost
pressure is fixed using an iterative technique which ensures  a smooth
pressure gradient in these regions.) The predicted  temperature
profile is rebinned to the same binning as the projected/deprojected 
spectra and compared with the observed spectral deprojection 
results. The $\chi^2$ difference between the observed and predicted 
temperature profiles is then calculated. The parameters for the mass model 
are stepped through a regular grid of values in the $r_{\rm
s}$-$\sigma$  plane to determine the best-fit values and 68 per cent
confidence limits.  Spherical symmetry and hydrostatic equilibrium are
assumed throughout.} with distances calculated using the code of 
Kayser, Helbig \& Schramm (1997). We have parameterized the mass 
profiles using a Navarro, Frenk \& White (1997; hereafter NFW) model
with  $\rho(r) = \rho_{\rm c}(z) \delta_{\rm c} / [({r/r_{\rm s}})
\left(1+{r/r_{\rm s}} \right)^2]$, where $\rho(r)$ is the mass density,
$\rho_{\rm c}(z) = 3H(z)^2/ 8 \pi G$ is  the critical density for
closure at redshift $z$, and $\delta_{\rm c} = {200 c^3
/ 3 \left[ {{\rm ln}(1+c)-{c/(1+c)}}\right]}$.  The normalizations of
the mass profiles may also be expressed in terms of  an equivalent
velocity dispersion, $\sigma = \sqrt{50} r_{\rm s} c H(z)$  (with
$r_{\rm s}$ in units of Mpc). The best-fit NFW model parameter values
and 68 per cent confidence limits are summarized in Table
\ref{table:nfw}.

In determining the results on the virial properties, 
we adopt $\Delta = 2500$, since $r_{2500}$ is well-matched to the 
outermost radii at which reliable temperature measurements can be made 
from the Chandra S3 data. (The $r_{2500}$ values for the NFW models are 
determined numerically, with confidence limits calculated using the 
$\chi^2$ grids. Note that $r_{2500}$ varies from $0.26-0.33\, r_{200}$ for 
the clusters in the present sample). We define $kT_{2500}$, the mean 
gas mass-weighted temperature within $r_{2500}$, as 
$kT_{2500} = \sum_{i=1}^{n} m_{\rm gas,i} kT_i/\sum_{i=1}^{n} m_{\rm gas, i}$ 
where $m_{\rm gas,i}$ and $kT_i$ are the gas mass and temperature (in keV)
in each radial shell for which an independent spectral determination 
of the temperature is made. The outer radius of shell $n$ is set 
to be equal to $r_{2500}$. Similarly, 
we define $L_{2500} = \sum_{i=1}^{n} L_i$, where $L_i$ is the bolometric 
luminosity in each radial shell. The best-fit values and 68 per cent 
confidence limits for $r_{2500}$, $M_{2500}, kT_{2500}$ and $L_{2500}$ 
are summarized in Table \ref{table:rmtl}.

\begin{table*}
\begin{center}
\caption{The total masses ($M_{2500}$, in units of $10^{14}$\Msun), 
mean gas mass-weighted temperatures ($kT_{2500}$, in keV) and bolometric 
luminosities ($L_{2500}$, in units of $10^{45}$\ergps) within radii 
$r_{2500}$ (in Mpc).} \label{table:rmtl}
\vskip -0.0truein
\begin{tabular}{ c c c c c c c c c c c c c }
\multicolumn{1}{c}{} &
\multicolumn{1}{c}{} &
\multicolumn{5}{c}{SCDM} &
\multicolumn{1}{c}{} &
\multicolumn{5}{c}{$\Lambda$CDM} \\
              && $E(z)$ &  $r_{2500}$            &       $M_{2500}$       &      $kT_{2500}$        &         $L_{2500}$         & ~ & $E(z)$ &      $r_{2500}$        &     $M_{2500}$         &      $kT_{2500}$        &         $L_{2500}$            \\
\hline	        	                             								         		 	                            
PKS0745-191   && 1.158  & $0.85^{+0.04}_{-0.05}$ & $6.06^{+0.69}_{-1.10}$ & $ 9.56^{+1.06}_{-0.75}$ &  $7.35^{+0.23}_{-0.34}$   & ~ & 1.050  & $0.68^{+0.03}_{-0.03}$ & $4.96^{+0.58}_{-0.68}$ & $ 9.55^{+1.06}_{-0.75}$ &  $4.28^{+0.11}_{-0.14}$      \\
Abell 2390    && 1.364  & $0.69^{+0.14}_{-0.09}$ & $4.41^{+3.22}_{-1.50}$ & $11.02^{+4.62}_{-1.83}$ &  $6.13^{+0.99}_{-0.77}$   & ~ & 1.122  & $0.64^{+0.15}_{-0.09}$ & $4.72^{+4.17}_{-1.79}$ & $11.65^{+3.18}_{-2.45}$ &  $4.20^{+0.46}_{-0.50}$      \\
Abell 1835    && 1.401  & $0.72^{+0.05}_{-0.03}$ & $5.41^{+1.13}_{-0.76}$ & $11.05^{+1.81}_{-1.19}$ &  $8.50^{+0.30}_{-0.19}$   & ~ & 1.135  & $0.66^{+0.06}_{-0.02}$ & $5.23^{+1.51}_{-0.45}$ & $11.23^{+1.72}_{-1.03}$ &  $5.61^{+0.27}_{-0.03}$      \\
MS2137-2353   && 1.505  & $0.49^{+0.03}_{-0.01}$ & $1.95^{+0.36}_{-0.15}$ & $ 5.53^{+0.52}_{-0.41}$ &  $3.33^{+0.09}_{-0.04}$   & ~ & 1.174  & $0.46^{+0.02}_{-0.03}$ & $1.89^{+0.25}_{-0.31}$ & $ 5.56^{+0.46}_{-0.39}$ &  $2.27^{+0.04}_{-0.05}$      \\
RXJ1347-1145  && 1.748  & $0.72^{+0.10}_{-0.08}$ & $8.27^{+3.77}_{-2.32}$ & $16.05^{+5.30}_{-2.65}$ &  $18.4^{+1.2}_{-1.0}$     & ~ & 1.271  & $0.73^{+0.08}_{-0.09}$ & $8.95^{+3.37}_{-2.81}$ & $15.34^{+4.75}_{-2.23}$ &  $13.6^{+0.7}_{-0.8}$        \\
3C295         && 1.765  & $0.42^{+0.03}_{-0.03}$ & $1.63^{+0.39}_{-0.31}$ & $ 5.51^{+0.78}_{-0.67}$ &  $1.74^{+0.08}_{-0.08}$   & ~ & 1.279  & $0.41^{+0.04}_{-0.03}$ & $1.60^{+0.45}_{-0.33}$ & $ 5.61^{+0.78}_{-0.75}$ &  $1.30^{+0.05}_{-0.06}$      \\
&&&&&&&&&&&& \\
\hline
\end{tabular}
\end{center}
 \parbox {7in}
{}
\end{table*}

We note that the data for PKS0745-191 do not quite reach to 
$r_{2500}$ and for this cluster we have extrapolated $L_{2500}$ 
using a power-law fit to the luminosity data in the range $0.6-0.9 r_{2500}$. 
(We assume that the temperature remains constant beyond $0.9 r_{2500}$.) 
The lensing and X-ray mass results for the clusters in our sample are 
discussed in detail by Allen \etal (2001b; Abell 2390), 
Schmidt \etal (2001; Abell 1835), Allen \etal (2001c; RXJ1347-1145) and 
Allen \etal, in preparation (PKS0745-191; MS2137-2353; see also Wise \etal in 
preparation). Although independent 
confirmation of the 
X-ray mass results for 3C295 (Allen \etal 2001a) is not available, we 
include this cluster in the analysis of the temperature-luminosity relation 
since in other ways it appears similar to the other objects in the sample.

\section{Results}

\subsection{A universal temperature profile for relaxed clusters.}

\begin{figure}
\vspace{0.3cm}
\hbox{
\hspace{0.0cm}\psfig{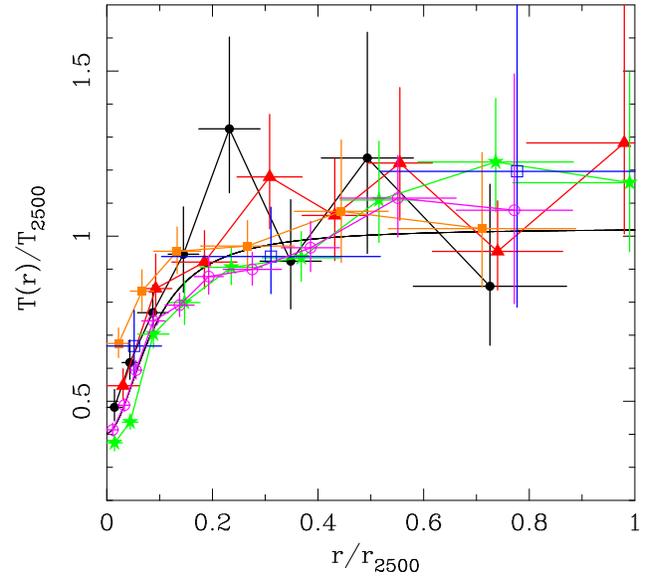}
}
\caption{The observed (projected) spectrally-determined temperature 
profiles in the clusters, scaled in units of $kT_{2500}$ and $r_{2500}$, 
for the $\Lambda$CDM cosmology. PKS0745-191: open circles, Abell 2390: dark
filled triangles, Abell 1835: grey filled stars, MS2137-2353: 
grey filled squares, RXJ1347-1145: dark filled 
circles, 3C295: open squares. The best-fit to the combined data set 
using the functional form in Section 3.1 is shown as 
the thin solid line.}\label{fig:kt}
\end{figure}

Fig. \ref{fig:kt} shows the observed (projected),
spectrally-determined  temperature profiles in the clusters, in units
of the mean gas  mass-weighted temperature, $kT_{2500}$ and  with the
radial axis scaled  in units of $r_{2500}$. The $\Lambda$CDM
cosmology is assumed. The clusters exhibit  similar
scaled-temperature profiles which rise within $r \sim 0.3 r_{2500}$  and
then remain approximately isothermal out to $r_{2500}$.
The combined data set can be modelled using a 
simple function of the form $T(r)/T_{2500} = 
T_0 + T_1[(x/x_{\rm c})^{\eta}/(1+(x/x_{\rm c})^{\eta}]$ 
where $x=r/r_{2500}$, $T_0 = 0.40\pm0.02$, 
$T_1=0.61\pm0.07$, $x_{\rm c}=0.087\pm0.011$ and $\eta=1.9\pm0.4$.  

\subsection{The mass-temperature relation}

\begin{figure*}
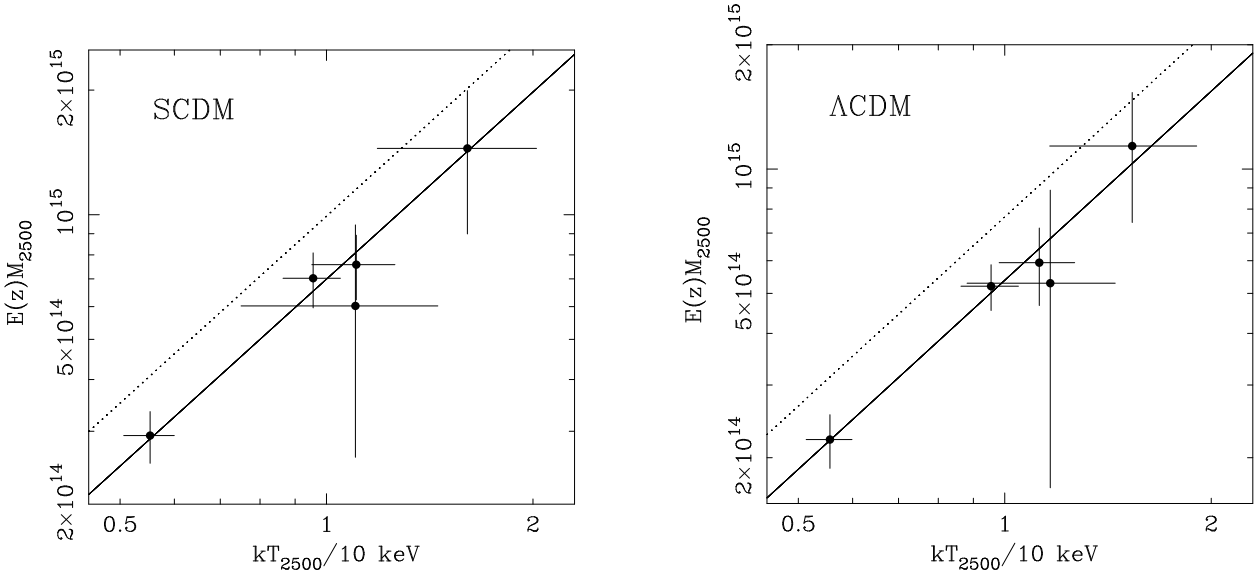

\vspace{0.2cm}
\hbox{
\hspace{0.2cm}\psfig{figure=mt_h50m1l0_final_label_log.ps,width=0.43 \textwidth,angle=270}
\hspace{1.3cm}\psfig{figure=mt_h70m3l7_final_label_log.ps,width=0.43 \textwidth,angle=270}
}
\caption{(a) The observed mass-temperature relation for the SCDM 
cosmology with $M_{2500}$ in \Msun~and $kT_{2500}$ in keV. The solid line 
is the best-fitting power-law model $E(z)M_{2500} = A [kT_{2500}/10]^{\alpha}$ 
with $A=6.99\pm0.57 \times 10^{14}$ and $\alpha=1.5$ (fixed). The dotted curve 
is the predicted result from the hydrodynamical simulations of Evrard \etal 
(1996). (b) The results for the $\Lambda$CDM cosmology. The solid line is the 
best-fitting power-law model with $A=5.38\pm0.74 \times 10^{14}$ and 
$\alpha=1.52\pm0.36$. The dotted curve is the predicted result from the 
hydrodynamical simulations of Mathiesen \& Evrard (2001).}\label{fig:mt}
\end{figure*}

Fig. \ref{fig:mt} shows the $M_{2500}-kT_{2500}$ relations for the
SCDM and $\Lambda$CDM cosmologies. Fitting only the data  for those
clusters for which independent confirmation of the X-ray mass  results
is available from lensing studies ({\it i.e.} excluding 3C295  from
the present sample) using a power-law model of the form 

\begin{equation}
E(z) \left( \frac{M_{2500}}{1 \,\Msun} \right) = A { \left( \frac{kT_{2500}}
{{\rm 10\, keV}} \right) }^\alpha
\end{equation} 

\noindent and a $\chi^2$ estimator, we obtain 
$A=6.91\pm0.68 \times 10^{14}$, $\alpha=1.43\pm0.26$ for SCDM 
($h=0.5$) and $A=5.38\pm0.52 
\times 10^{14}$, $\alpha=1.51\pm0.27$ for $\Lambda$CDM ($h=0.7$).
Fixing the slope at $\alpha=1.5$, we measure normalizations of 
$A=6.99\pm0.60\times 10^{14}$ (SCDM) and $A=5.38\pm0.46\times 10^{14}$ 
($\Lambda$CDM). Uncertainties are quoted at the 68 per cent 
confidence level.

Using the modified least-squares estimator of Fasano \& Vio (1988), 
which accounts for errors in both axes, we obtain values 
$A=7.02\pm 0.98 \times 10^{14}$, $\alpha=1.47\pm0.36$ for SCDM and
$A=5.38\pm0.74 \times 10^{14}$, $\alpha=1.52\pm0.36$ for the 
$\Lambda$CDM cosmology. 
We conclude that the measured slope is 
consistent with the expected value of $\alpha=1.5$ in all cases.

The dotted curves in Figs. \ref{fig:mt}a,b show the predicted
(zero-redshift) relations for the SCDM [$E(z)M_{2500} = 9.9\pm1.5
\times 10^{14} kT^{1.5}$] and $\Lambda$CDM [$E(z)M_{2500} = 7.7\pm0.6
\times 10^{14} kT^{1.52\pm0.03}$] cosmologies from the hydrodynamical
simulations of Evrard \etal (1996) and Mathiesen \& Evrard (2001),
respectively. We have scaled the predicted curves from $\Delta=500$ to
$\Delta=2500$ assuming $M_{2500} =  M_{500} (2500/500)^{-0.5} 
(T_{2500}/T_{500})^{1.5}$, which is consistent with the range of
best-fit NFW mass models, and the SCDM simulations of Evrard \etal
(1996; see their Table 5). Note that allowing for the presence of
temperature gradients ($T_{2500} \neq T_{500}$) in the simulated
clusters when applying this scaling does not affect the best-fit
parameters for the theoretical $M_{2500}-kT_{2500}$ curves, since the
curves are simply mapped onto themselves. For both SCDM and
$\Lambda$CDM,  the predicted normalization lies approximately 40 per
cent above the observed  value.

\subsection{The temperature-luminosity relation}

\begin{figure}
\vspace{0.3cm}
\hbox{
\hspace{0.2cm}\psfig{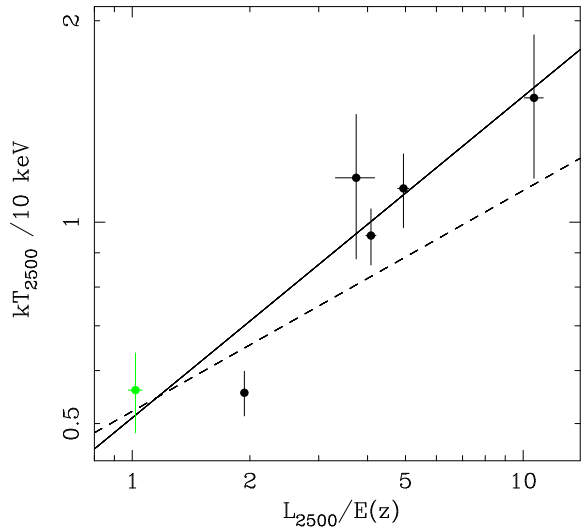}
}
\caption{The observed temperature-luminosity relation for the 
$\Lambda$CDM cosmology with $kT_{2500}$ in keV and $L_{2500}$ in units 
of $10^{45}$ \ergps. The solid line is the best-fitting power-law model 
of the form $kT_{2500}/10 = B {[L_{2500}/10^{45} E(z)]}^{\beta}$ using 
the BCES 
estimator, for which $B=0.51\pm0.05$ and $\beta=0.48\pm0.06$. The 
data for 3C295 are in lighter shading. The dashed line
shows the best-fit curve with $\beta=0.33$ fixed which provides a poor 
description of the data.}\label{fig:lt}
\end{figure}

Fig. \ref{fig:lt} shows the $kT_{2500}-L_{2500}$ relation for 
the $\Lambda$CDM cosmology. Fitting the $kT_{2500}-L_{2500}$ 
data for all six clusters using a power-law model of the form 

\begin{equation}
\left( \frac{kT_{2500}} {{\rm 10\, keV}} \right) = B  {\left( \frac{L_{2500}}{{\rm 10^{45}\ergps}E(z)} \right)}^{\beta}
\end{equation} 

\noindent and a $\chi^2$ estimator, we obtain 
$B=0.43\pm0.05$, $\beta=0.45\pm0.09$ for SCDM, and  
$B=0.42\pm0.05$, $\beta=0.56\pm0.10$ for $\Lambda$CDM. 
Using the BCES($X_2|X_1$) estimator of Akritas \& Bershady (1996), which 
accounts for errors in both axes and the 
presence of possible intrinsic scatter, we obtain $B=0.48 \pm 0.07$, 
$\beta=0.46 \pm0.08$ for SCDM and 
$B=0.51 \pm 0.05$, $\beta=0.48\pm0.06$ for $\Lambda$CDM. 

We conclude that the slope of the temperature-luminosity 
relation for the present sample of hot, relaxed clusters is  
consistent with the predicted value of $\beta=0.5$ (Section 1). Fixing
$\beta=0.33$ results in a poor fit: $\chi^2=12.5$ for 5 degrees 
of freedom, as opposed to $\chi^2=6.7$ with $\beta=0.5$ ($\Lambda$CDM).

\section{Discussion}

We have shown that within radii $r_{2500}$, corresponding to a fixed
density contrast $\Delta =2500$ with respect to the critical density
at the redshifts of the clusters, the temperature profiles for the 
present sample of luminous, relatively relaxed lensing clusters 
exhibit an approximately universal form which rises
within $r \sim 0.3\, r_{2500}$ and then remains  approximately constant
out to $r_{2500}$. The enclosed masses, bolometric
luminosities and mean gas mass-weighted temperatures within these radii 
scale in manner consistent with the predictions from the simple virial
relations outlined in Section 1. We have confirmed the presence of a 
systematic offset of $\sim 40$ per cent between the normalizations of the
observed and predicted $M_{2500}-kT_{2500}$ curves, in the sense that
the predicted temperatures are too low for a given mass, for both the
SCDM and $\Lambda$CDM cosmologies.
 
An important aspect of the present study is that independent confirmation 
of the X-ray mass measurements is available from gravitational lensing
studies. For both Abell 2390 and RXJ1347-1145, the X-ray and weak
lensing mass profiles are consistent within their 68 per cent 
confidence limits. For Abell 1835, 2390, MS2137-2353 and PKS0745-191, the 
observed strong lensing configurations (on scales $r \sim 20-80\,h^{-1}$kpc) 
can be explained by mass models
within the 68 per cent Chandra confidence contours, although redshift
measurements for the arcs  (which are required to define the lensing
masses precisely) are not available in all  cases.\footnote{For
RXJ1347-1145, a two-component mass  model, consistent with the complex
X-ray structure observed in the   southeast quadrant, is required to
explain the strong lensing data.}  Thus, the presence of significant 
non-thermal pressure support (\eg arising from turbulent and/or bulk 
motions and/or magnetic fields) can be excluded. We conclude that the 
systematic uncertainties associated with the individual mass measurements 
are small ($<20$ per cent).

The offset between the observed and simulated mass-temperature curves
cannot be explained by invoking an earlier formation redshift for the
observed clusters (we assume that the clusters form at the redshifts
they are observed) since, for the measured NFW mass distributions,
$M_{2500}(z)$ drops as fast or faster than $E(z)$ rises as the
formation redshift is increased. Our results suggest that on 
the spatial scales studied here, important physics may be missing from 
the reference simulations.  One possible candidate is radiative cooling of the
X-ray gas, which the Chandra data show to be significant within  $r
\sim 0.2 r_{2500}$ (\eg Allen \etal 2001a,b,c; David \etal 2001; 
Schmidt \etal 2001).
Pearce \etal (2000) show that the introduction  of radiative cooling
into their hydrodynamical simulations can lead to  central temperature
drops similar to those in Fig. \ref{fig:kt}. These authors also argue 
that cooling can lead to a significant increase in the mass-weighted
temperature within $r \sim r_{2500}$ (as cooled, low-entropy gas is
deposited and warmer, high-entropy material flows inwards and is 
compressed), which may be sufficient  to account for the discrepancy
between the observed and simulated curves.  Detailed simulations of
the $M_{2500}-kT_{2500}$ relation for large a  sample of massive
clusters, including the effects of radiative cooling,  are required to
address this issue.

The results presented in this paper should provide a useful calibrator
for future studies of the X-ray properties of galaxy clusters.  In
future work we will examine the constraints that the present data
place on radial variations in the X-ray gas mass fraction in the
clusters and, therefore, $\Omega_{\rm m}$. We will also explore the
ability of  different parameterized mass models to explain the
observed X-ray gas temperature and density profiles.

\vspace{0.3cm}

\noindent SWA and ACF acknowledge the support of the Royal Society. 

\end{document}